\documentclass{article}
\usepackage{mlspconf,amsmath,graphicx}
\usepackage{amssymb}
\usepackage[table]{xcolor}
\usepackage{color}
\usepackage{balance}
\usepackage{cite}
\usepackage{booktabs}
\usepackage{multirow}
\usepackage[ruled,vlined]{algorithm2e}

\usepackage{hyperref}
\hypersetup{
    colorlinks=true,
    linkcolor=blue,
    filecolor=magenta,      
    urlcolor=cyan,
}

\let\OLDthebibliography\thebibliography
\renewcommand\thebibliography[1]{
  \OLDthebibliography{#1}
  \setlength{\parskip}{0pt}
  \setlength{\itemsep}{3pt plus 0.3ex}
}

\copyrightnotice{978-1-7281-6662-9/20/\$31.00 {\copyright}2020 IEEE}

\toappear{2020 IEEE International Workshop on Machine Learning for Signal Processing, Sept.\ 21--24, 2020, Espoo, Finland}

\usepackage[]{algorithm2e}
\usepackage{wasysym}
\usepackage{bm}
\usepackage{url,cite}
\usepackage{enumitem} 
\usepackage{pifont}

\newcolumntype{?}{!{\vrule width 1pt}}
\title{AutoClip: Adaptive Gradient Clipping for Source Separation Networks}
\name{Prem Seetharaman$^{1}$\sthanks{This work has made use Mystic (CRI-1730689) at IllinoisTech.}, Gordon Wichern$^2$, Bryan Pardo$^1$, Jonathan Le Roux$^2$}
\address{
$^1$Northwestern University, Evanston, IL, USA\\
$^2$Mitsubishi Electric Research Laboratories (MERL), Cambridge, MA, USA\\
}

\begin{document}
%
\maketitle
\begin{abstract}
Clipping the gradient is a known approach to improving gradient descent, but requires hand selection of a clipping threshold hyperparameter. We present AutoClip, a simple method for automatically and adaptively choosing a gradient clipping threshold, based on the history of gradient norms observed during training. Experimental results show that applying AutoClip results in improved generalization performance for audio source separation networks. Observation of the training dynamics of a separation network trained with and without AutoClip show that AutoClip guides optimization into smoother parts of the loss landscape. AutoClip is very simple to implement and can be integrated readily into a variety of applications across multiple domains.
\end{abstract}
\section{Introduction}
\label{sec:intro}
Audio source separation, a central task in computer audition, is the study of how to break down complex auditory scenes into their constituent parts (e.g., isolating the voice of a single speaker in a crowd). Robust audio source separation has many practical applications, such as video conferencing, hearing aids, speech enhancement, and home voice assistants. In recent years, deep neural networks have greatly advanced the state of the art in source separation across multiple audio domains, such as speech, music, and environmental sounds. However, optimizing deep networks for audio source separation remains a tricky endeavor, requiring careful tuning of hyperparameters and choice of optimizer, network architecture, and loss function.

The loss landscape on which a neural network is optimized is often non-smooth and filled with local minima. This is especially true in the case of recurrent neural networks, which are vulnerable to both exploding and vanishing gradient issues \cite{bengio1994learning}. Gradient clipping \cite{pascanu2013difficulty, mikolov2012statistical, zhang2019gradient,sundermeyer2012lstm} attempts to resolve the former issue: exploding gradients. Gradient clipping has been found to be a highly effective and necessary ingredient for state-of-the-art performance in applications such as language modeling \cite{mikolov2012statistical, sundermeyer2012lstm}.
It has also proved crucial to the optimization process of recent state-of-the-art source separation algorithms~\cite{luo2019convTasNet, wichern2019wham, zeghidour2020wavesplit}. In gradient clipping, if the norm of the gradient vector (taken over all parameters) exceeds the clipping value, the gradient vector is scaled so that its norm does not exceed that value. The clipping value is typically set by hand. 

The gradient clipping hyperparameter is very sensitive to the loss function, the scaling of the data, and the network architecture. As an illustrative example, consider two implementations of the same mask inference architecture \cite{Wang2018ICASSP04Alternative}. Mask inference networks are typically applied to audio preprocessed with a short term Fourier transform (STFT). Assume that, for one network, the STFT normalizes each window, and for the other, the STFT does not normalize. This can result in data scales that differ by orders of magnitude. One cannot specify a single gradient clipping value appropriate for both, otherwise identical, implementations\footnote{This situation has occurred in practice, resulting in degraded performance after changing STFT functions with the same clipping threshold.}. This is in contrast to other facets of the optimization process, such as using an Adam optimizer \cite{Kingma2015Adam}, which adapts to the statistics of the gradients. 

In this work, we propose AutoClip, a simple adaptive gradient clipping procedure which removes the need to hand-tune the clipping parameter, transfers easily across multiple loss functions, and is scale-invariant by design. In experiments, we show the impact of AutoClip on the optimization of source separation networks. Our experiments also provide empirical evidence for the need for gradient clipping when training source separation networks. An implementation is provided here: \href{https://github.com/pseeth/autoclip}{https://github.com/pseeth/autoclip}.

\section{Gradient Clipping}

Given a function $f(X;\theta)$ computed on data $X$ and parameterized by $\theta$, and a learning rate $\lambda$, 
a gradient descent update at iteration $t$ of the current parameters $\theta_{t-1}$ to $\theta_{t}$ is defined as:
\begin{equation}
    \theta_{t} = \theta_{t-1} - \lambda \nabla_{\theta} f(X;\theta_{t-1}).
\end{equation}
Note that a stochastic version of this update is more commonly used, with $\nabla_{\theta} f(X_t;\theta_{t-1})$ being computed on minibatches $X_t$. Gradient clipping enforces an upper bound on the update of $\theta$, by placing a max on the norm of the gradient:
\begin{align}
    \label{eq:grad_clip}
    \theta_{t} &= \theta_{t-1} - \lambda h_c \nabla_{\theta} f(X_t;\theta_{t-1}), \\
    h_c &= \text{min}\{\frac{\eta_c}{||\nabla_{\theta} f(X_t;\theta_{t-1})||}, 1 \}.
\end{align}
Here, $\eta_c$ is a clipping value hyperparameter that needs to be carefully chosen by the end user. Note that this clipping scheme is the so-called clip by norm, not clip by value, where individual values of the gradient vectors are clipped if they go beyond a pre-set value. In clip by norm, the entire gradient is scaled if the norm of the gradient exceeds a threshold. In stochastic gradient descent (SGD), this places a maximum on the step size that can be taken during training, preventing the optimization from going too far in the direction of a gradient with very large magnitude. If the gradient is below the threshold, the optimization is unaffected. 

The reason why gradient clipping stabilizes and accelerates training of neural networks (especially recurrent ones) is an active area of research, with both empirical work showing its efficacy \cite{pascanu2013difficulty} as well as theoretical work analyzing the dynamics of SGD with gradient clipping \cite{zhang2019gradient}.

In gradient clipping, selection of $\eta_c$ is very important. If it is set too high, then the gradient norm will always be smaller than that, and clipping is never applied. If set too low, then the step size taken by the network may be too small. In practice, a heuristic proposed by Pascanu et al.~\cite{pascanu2013difficulty} is often used for setting the clipping threshold. First, start training and observe the gradient norms of each batch for a sufficient number of updates. A set of reasonable values can then be selected to serve as adequate settings for $\eta_c$, and an optimal value determined by cross-validation, or, as a simpler alternative, a value anywhere between 5x and 10x the average gradient norm that was observed can be used. This ad-hoc method must be done for each network, loss function, and dataset. There is no ``one-size-fits-all'' number for the gradient clipping threshold.

\section{AutoClip}
AutoClip is an automated and dynamic approach to setting the clipping value hyperparameter $\eta_c$ based on the statistics of the history of the gradient norms observed while training. 
AutoClip sets $\eta_c(t)$ at each training iteration $t$, where each iteration corresponds to the processing of one minibatch. AutoClip keeps track of the gradient norm on every batch seen during training. Given the history of the gradient norms $G_h(t)$ up to iteration $t$, it calculates the gradient norm that would reach the $p$-th percentile value $n_p(t)$ of the current history. It then sets $\eta_c(t) = n_p(t)$. This is shown in Algorithm~\ref{alg:AutoClip}.

While AutoClip is agnostic to the optimizer and can be used with SGD, in our work we use gradient clipping in conjunction with the Adam optimizer, rather than SGD. Adam has become the de facto optimizer for source separation networks \cite{luo2019convTasNet, leroux2019phasebook, LeRoux2018SISDR, zeghidour2020wavesplit, Wang2018ICASSP04Alternative} and is thus of particular interest to this work. The standard Adam equations \cite{Kingma2015Adam} for updating the parameters of the network $\theta_{t-1}$ are used, with the only difference that the gradients are replaced by their (potentially) clipped versions before the Adam optimizer updates the estimate of the first ($m_{t}$) and second ($v_{t}$) moments of the gradients.


\begin{algorithm}
\SetAlgoLined
\SetKwInOut{Input}{Input}
\SetKwInOut{Output}{Output}
\Input{
    $p$ -- Percentile cutoff for setting $\eta_c(t)$.\\
    $\theta_0$ -- Initial network parameters.\\
    $D$ -- Data to take batches from.\\
    $T$ -- Total number of iterations.
}
 $G_h(0) \gets [\,]$\;
 \ForEach{iteration $t \in (1,\dots, T)$}{
  Calculate loss on batch of data $X_t \in D$\;
  $G_h(t) = G_h(t-1)$.append($\nabla_{\theta} f(X_t; \theta_{t-1})$)\;
  $n_p(t) \gets p$-th percentile of $G_h(t)$\;
  $\eta_c(t) \gets n_p(t)$\;
  Clip gradients using $\eta_c(t)$\;
  $\theta_{t} \gets$ Step optimizer on network parameters
 }
 \caption{Training with AutoClip}\label{alg:AutoClip}
\end{algorithm}

The dynamics of AutoClip lead to an adaptive setting of $\eta_c(t)$ that is determined dynamically by the data, the network, and the loss, as opposed to one selected carefully by a user after observing the training dynamics. 
A user only has to specify what percentile to clip to. As mini-batch stochastic optimization is prone to outliers, using robust statistics \cite{huber2004robust} via percentiles would mitigate this issue. This setting can be transferred across multiple optimization scenarios, as it is defined relative to the data and loss, rather than an absolute value that is sensitive to these factors as well as to implementation.

For higher settings of $p$, less clipping is applied to the gradients. If $p = 100$, then no clipping is applied during optimization. If $p = 0$, then every gradient is clipped to the minimum gradient seen during training until then. For low values of $p$, the clip value $\eta_c(t)$ will have more ``inertia,'' as it will not update significantly without a long sequence of high gradient norms. For higher values of $p$, $\eta_c(t)$ will be much more responsive to seeing higher gradient norms. As a result, AutoClip will only raise the clipping value during optimization if a long-enough sequence of high gradient norms justifies it.

\subsection{Relationship to other optimization approaches}
Much recent work in optimization has centered around ``denoising'' the gradient during training. AdaGrad\cite{duchi2011adaptive} achieves this by keeping track of the sum of the squared gradients for each parameter. One drawback of AdaGrad is that the learning rate will decay to an infinitesimally small number as more iterations are taken. A refinement of AdaGrad, AdaDelta \cite{zeiler2012adadelta}, resolves this issue by accumulating the sum of the squared gradients over a window. Adam \cite{Kingma2015Adam} addresses outliers in the gradient by keeping track of the first and second moments of the gradient for each parameter using an exponential moving average. After correcting for bias, the optimizer step taken on each parameter is adjusted based on the observed variance of the gradient history. Adam can be seen as combining the best of AdaGrad with RMSProp \cite{Tieleman2012}. AutoClip can be applied in conjunction with any existing optimization approach (including AdaGrad, Adam, SGD, etc.). AutoClip denoises the gradients via adaptive gradient clipping, which is done prior to the optimization step. 

For $p=0$, assuming that a small gradient norm appears early in training, all subsequent gradients will be normalized to that constant value, so all clipped gradients passed to the optimizer will have the same norm. For a scale-invariant optimizer such as Adam, such a setting is equivalent to using normalized gradients, which was shown to improve optimization~\cite{yu2017block}.

\section{Experimental design}
Our experiments are designed to investigate the impact of AutoClip on optimization. Our primary research question is whether a single setting of AutoClip can transfer easily across loss functions that have vastly different scales. When setting gradient clipping manually, the thresholds must be carefully chosen based on observing the gradient norms of training for every network and loss function individually. AutoClip scales the clipping threshold automatically, potentially enabling a ``set-and-forget'' approach to applying gradient clipping.

Our secondary research question stems from the fact that AutoClip is interpretable, because it is relative to the actual gradient norms. For example, if $p=10$, then we know $90$\% of the gradients are typically being clipped during training. By varying $p$, we can draw conclusions about the impact of more or less aggressive clipping during optimization.

We apply AutoClip to the problem of separating individual speech streams from a mixture of concurrent speech. We use a standard dataset in the speech separation literature, WSJ0-2mix \cite{Hershey2016}, which consists of 20,000 2-speaker mixtures for training, 5,000 for validation, and 3,000 for test. The sample rate of all audio files was fixed to be 8 kHz. We use 32 ms windows, 8 ms hop length, and the square root of the Hann window as our window function for computing the STFT.

\subsection{Loss functions}

We investigate the interaction between AutoClip and five commonly used loss functions in the source separation literature. 
Because these loss functions have different scales, their gradients will also likely have different scales\footnote{$\nabla (a f) = a \nabla f$, where $a$ is a constant.}. These gradients will thus have different norms, and therefore the optimal gradient clipping value $\eta_c$ will vary. 

We chose the classic deep clustering loss ($\mathcal{L}_{\text{DC}}$) \cite{Hershey2016}, the whitened k-means loss ($\mathcal{L}_{\text{WKM}}$) \cite{Wang2018ICASSP04Alternative}, a mask inference loss based on the $L_1$ distance between the estimated source and the actual source ($\mathcal{L}_{\text{MI}}$) \cite{Wang2018ICASSP04Alternative}, a Chimera multi-task loss function that combines $\mathcal{L}_{\text{MI}}$ and $\mathcal{L}_{\text{WKM}}$ ($\mathcal{L}_{\text{MI+WKM}}$) \cite{Luo2017,Wang2018ICASSP04Alternative}, and a waveform loss where the audio output of the network is optimized via the signal-to-noise ratio ($\mathcal{L}_{\text{SNR}}$) \cite{kavalerov2019universal}. 

The mask inference loss uses the truncated phase-sensitive spectrum approximation (PSA) loss \cite{Erdogan2015} to compare an estimated spectrogram $\hat{S}$ with the ground truth spectrogram $S$:
\begin{equation}
\mathcal{L}_{\text{MI}} = \frac{1}{N} \Big\| |\hat{S}| - \operatorname{T}_{0}^{|X|}\left(|S| \circ \cos(\theta_{\hat{S}} - \theta_{S})\right) \Big\|_1, 
\end{equation}
where $N$ denotes the number of time-frequency points in $S$, $X$ the input mixture, $|Y|$ and $\theta_{Y}$ the magnitude and phase of a spectrogram $Y$, and $\operatorname{T}_{a}^{b}(x)= \min(\max(x,a),b)$ denotes truncation to the range $[a,b]$.
The values of $\mathcal{L}_{\text{MI}}$ are on the order of 1e-5. The deep clustering loss compares the affinity matrix of embeddings $V$ for all time-frequency (TF) points with that of ground truth assignments $Y$, introducing weights $W$ for every TF point:
\begin{equation}
\mathcal{L}_{\text{DC}} = ||W^{1/2}(VV^T - YY^T) W^{1/2}||^2_F.
\end{equation}
The values of $\mathcal{L}_{\text{DC}}$ range between $0$ and $1$, when $W$ is normalized for the number of TF points. The whitened k-means loss is a self-normalizing variant of $\mathcal{L}_{\text{DC}}$:
\begin{equation}
\mathcal{L}_{\text{WKM}} = D - \text{tr}((V^TV)^{-1} V^TY (Y^TY)^{-1} Y^TV)
\end{equation}
where $D$ is the embedding size. The range of $\mathcal{L}_{\text{WKM}}$ is between $0$ and $D$, where $D$ is typically around $20$. 

The multi-task loss function -- the Chimera loss in \cite{Luo2017,Wang2018ICASSP04Alternative} -- combines the mask inference loss with the whitened k-means loss, weighting each using a constant factor $\alpha$:
\begin{equation}
\mathcal{L}_{\text{MI+WKM}} = \alpha \mathcal{L}_{\text{MI}} + (1 - \alpha) \mathcal{L}_{\text{WKM}}.
\end{equation}
In this work, $\alpha$ is chosen to be $0.75$. Because of the large difference between the magnitudes of $\mathcal{L}_{\text{MI}}$ and $\mathcal{L}_{\text{WKM}}$ (in practice orders of magnitude apart), this results in $\mathcal{L}_{\text{WKM}}$ mostly dominating the optimization. The range of $\mathcal{L}_{\text{MI+WKM}}$ is between $0$ and $(1 - \alpha) D$ (typically between $0$ and $5$).

The signal-to-noise ratio compares the time-domain audio estimate $\hat{s}$ and the time-domain ground truth source $s$ \cite{kavalerov2019universal}: 
\begin{equation}
\mathcal{L}_{\text{SNR}} = - 10 \log_{10} \left(\frac{||s||^2}{||s-\hat{s}||^2}\right).
\end{equation}
$\mathcal{L}_{\text{SNR}}$ is the negative SNR so that it is minimized during optimization. It can range anywhere between $-\infty$ and $+\infty$, but more typically stays between $-20$ and $20$.

\subsection{Network architectures}

We train identical networks for each of these loss functions, with minor differences to account for their losses' particularities. At the core of each network is a stack of 4 bidirectional LSTM layers with 600 hidden units in each direction. The input to each network is the log-magnitude spectrogram of the mixture. The networks that are trained with $\mathcal{L}_{\text{DC}}$ and $\mathcal{L}_{\text{WKM}}$ output 20-dimensional embeddings for each time-frequency point. These embeddings are unit-norm with sigmoid activation. The mask inference network outputs two masks with sigmoid activation. The Chimera network trained with $\mathcal{L}_{\text{MI+WKM}}$ outputs an embedding and two masks. The network that is trained with $\mathcal{L}_{\text{SNR}}$ does so by using an inverse STFT within the network: it outputs two masks which are applied to the magnitude spectrogram; the masked STFT is then inverted using the mixture phase to the time-domain audio for each source within the network. We use permutation-invariant versions of $\mathcal{L}_{\text{SNR}}$, $\mathcal{L}_{\text{MI}}$, and $\mathcal{L}_{\text{MI+WKM}}$ \cite{Hershey2016,Kolbaek2017}.

\begin{table}[]
\centering
\begin{tabular}{c | ccccc}
\toprule
$p$ & $\mathcal{L}_{\text{DC}}$  & $\mathcal{L}_{\text{WKM}}$   & $\mathcal{L}_{\text{MI}}$      & $\mathcal{L}_{\text{MI+WKM}}$  & $\mathcal{L}_{\text{SNR}}$  \\
\midrule
\phantom{10}0       & 10.7 &  11.1  & 10.0 &  11.2  & \phantom{1}9.9  \\ \midrule
\phantom{10}1       & 10.7 &  11.2     & 10.3  &  11.3  & 10.2   \\ 
\phantom{1}10     & 10.8 &  11.0      & 10.2 &  11.3      &  10.4 \\ 
\phantom{1}25     & 10.7 &   11.0     & \phantom{1}9.9  &  11.3      &  10.3  \\
\phantom{1}50     & 10.7 & 11.0 & \phantom{1}9.2  &  11.2      &  \phantom{1}9.9   \\ 
\phantom{1}90      & 10.5 &  11.0      & \phantom{1}8.7  &  11.1   & \phantom{1}9.5    \\ \midrule
100     & 10.2 &  10.8      & \phantom{1}8.5  &   10.9     &  \phantom{1}8.3   \\ 
\bottomrule
\end{tabular}
\caption{Performance in terms of SI-SDR [dB] on the WSJ0-2mix test set for each loss function when using AutoClip with varying percentile thresholds $p$. Higher values are better.}
\label{tab:AutoClip}
\end{table}

\subsection{Training and evaluation}
\label{sec:training}
All networks are trained with identical hyperparameters and are initialized with the same random seed. 
We use the Adam optimizer with an initial learning rate of 1e-3, and a sequence length of 400 frames (25536 samples for the network trained with $\mathcal{L}_{\text{SNR}}$) which are selected from random offsets within each utterance. Mixtures that are too short are padded with zeros to the required length. We use a batch size of 25 and train for 100 epochs. For each loss function, we experiment with the value of $p$, the percentile threshold in AutoClip, setting it to $0$, $1$, $10$, $25$, $50$, $90$, and $100$. Setting $p$ to $0$ corresponds to the most aggressive clipping strategy of ``min-clipping,'' where every gradient is clipped to the minimum gradient norm seen so far during training. Setting it to $100$ corresponds to no gradient clipping at all applied during training. 
We evaluate the performance of each network on the WSJ0-2mix test set using scale-invariant source-to-distortion ratio (SI-SDR) \cite{LeRoux2018SISDR}.

\begin{figure*}[h]
    \centering
    \includegraphics[width=\linewidth]{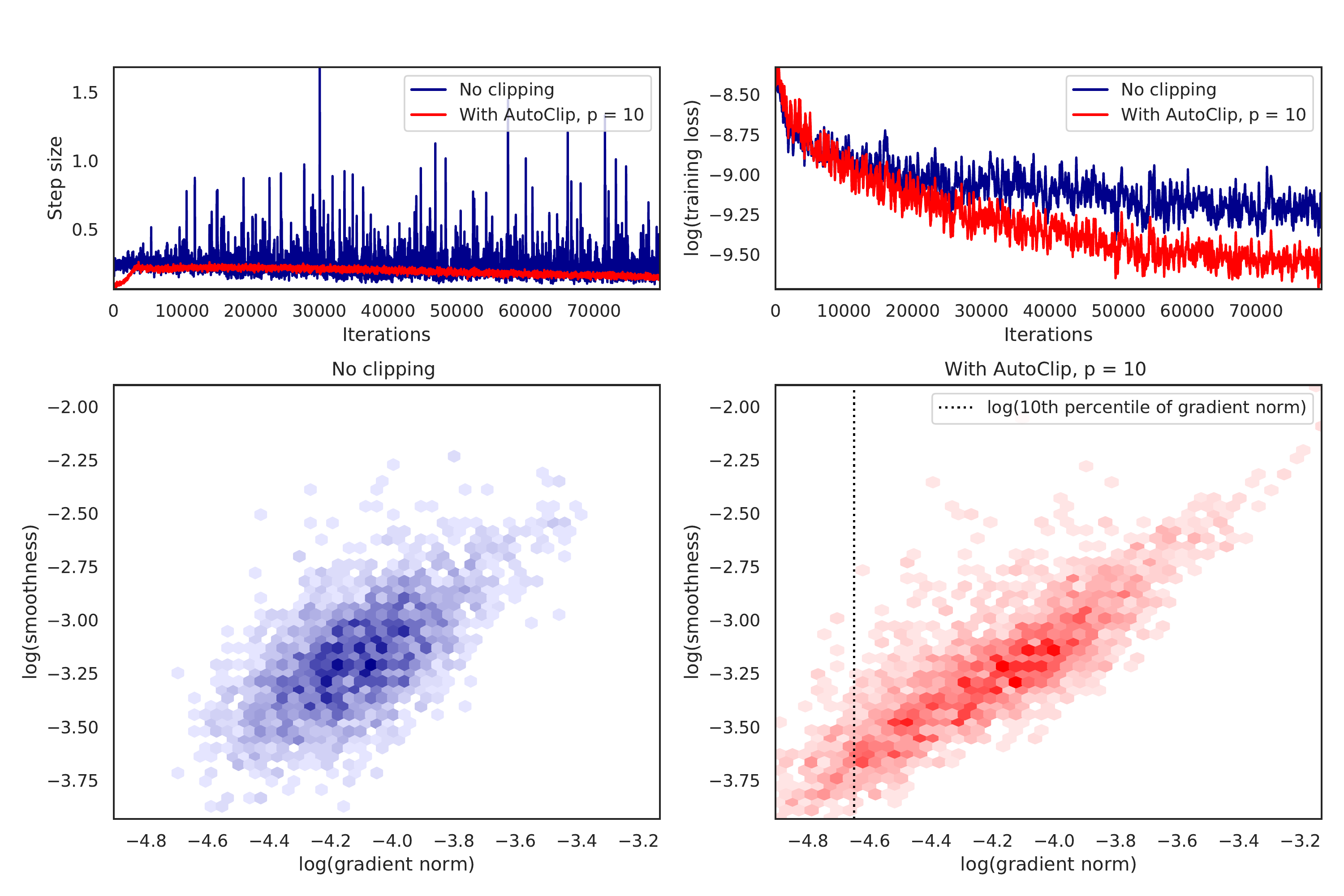}
    \caption{Training dynamics of a smaller mask inference network (2 BLSTM layers with 300 hidden units) with $\mathcal{L}_{\text{MI}}$, with and without AutoClip. The top left figure shows the norm of the step size taken on the model parameters. The top right figure shows the training loss over time, showing that AutoClip leads to better optimization. The bottom figures show the relationship between gradient norm and a measure of smoothness along the training trajectory. Statistics were recorded every 20 iterations during training.  With AutoClip, we observe a stronger correlation (r-value of $.86$), compared to without (r-value of $.62$). All gradients to the right of the dashed black line in the bottom right plot are clipped. We show the location of the AutoClip threshold at the end of training. The threshold changes during training, see an animated version: \href{https://imgur.com/a/rfbKUDE}{https://imgur.com/a/rfbKUDE}.}
    \label{fig:dynamics}
\end{figure*}

\section{Results and discussion}
The performance of each network we trained is shown in Table \ref{tab:AutoClip}. As expected, the bottom row, corresponding to $p=100$ where no gradient clipping is applied, gets consistently worse results across the board for every loss function. In the case of $\mathcal{L}_{\text{MI}}$ and $\mathcal{L}_{\text{SNR}}$, not applying clipping results in a huge performance drop of nearly $2$ dB. To the best of our knowledge, this is the first formal reporting of this phenomenon for source separation. Setting $p=1$ or $p=10$ results in vastly improved performance across the board for each loss function. As a reminder, setting a static clipping threshold for each of these loss functions would require individual hyperparameter search for each scenario. With AutoClip, one can set $p=10$  to get greatly improved performance, showing that AutoClip can be a ``set-and-forget'' approach to gradient clipping. Even continuously updating the clipping value to be very low ($p=1$) or even the minimum ($p=0$) of the gradients seem to greatly aid optimization.

Prior work optimizing $\mathcal{L}_{\text{DC}}$, $\mathcal{L}_{\text{MI}}$, and $\mathcal{L}_{\text{MI+WKM}}$ all used static clipping thresholds \cite{Wang2018ICASSP04Alternative}. For $\mathcal{L}_{\text{DC}}$, we obtain higher performance ($10.7$ dB) than reported in prior work ($10.2$ dB). We also observe this for $\mathcal{L}_{\text{MI}}$ ($10.3$ dB with AutoClip vs $10.0$ dB in prior work), $\mathcal{L}_{\text{WKM}}$ ($11.2$ dB with AutoClip vs $10.4$ dB in prior work), and $\mathcal{L}_{\text{MI+WKM}}$ ($11.3$ dB with AutoClip vs $11.2$ dB in prior work). AutoClip discovers optimal clipping thresholds, without requiring an exhaustive hyperparameter search.

The results in Table \ref{tab:AutoClip} hints at very different learning dynamics when training with and without AutoClip. To investigate this further, we observe the training dynamics of a smaller speaker separation network (as capturing detailed training behavior is computationally expensive). The smaller network has 2 BLSTM layers with 300 hidden units. The training recipe is identical to that in Section \ref{sec:training}. We compare training the network with AutoClip ($p = 10$) and without ($p = 100$). Every 20 iterations, we record the training loss, the step size, the gradient norm, and the local smoothness. Given model parameters $\theta_{t-1}$ and $\theta_{t}$ at consecutive time steps, the step size is computed as $||\theta_{t} - \theta_{t-1}||_2$, the $L^2$ norm of the difference in the model parameters. Local smoothness is measured via the local gradient Lipschitz constant, as used in prior work \cite{zhang2019gradient, santurkar2018does}. 

Fig.~\ref{fig:dynamics} shows that the step size varies more smoothly when using AutoClip. AutoClip induces an almost constant step size, which decays slowly over the course of training. Initially, the gradient norms are small, leading to small step sizes. As the gradient norms get bigger, so does the clipping threshold. This manifests as an initial ramp in the step size, similar to learning rate warmup, another popular training trick for deep networks \cite{liu2019radam}. After this initial stage, the step size then decays slowly with little variance. This corresponds to better optimization, as shown by the lower training loss, and higher test performance ($9.2$ dB vs $8.1$ dB).

In the two lower plots in Fig.~\ref{fig:dynamics}, we show the relationship between the gradient norm and the local smoothness. When training with AutoClip, there is a better correlation between the two. With AutoClip, areas of low gradient norm (e.g. minima) are also smoother.  Smoother minima are believed to result in better generalization performance \cite{keskar2016large, hochreiter1997flat}. AutoClip does not clip the gradient in relatively smooth regions.

\section{Conclusion}
We have presented AutoClip, a simple method for adaptively choosing a threshold for gradient clipping based on the history of gradient norms observed during training. AutoClip obviates the need for a hand-tuned clipping threshold and generalizes across loss functions with different scales. Experiments show that AutoClip results in better test performance for source separation networks. We examined the training dynamics of a separation network trained with and without AutoClip, showing that AutoClip stabilizes optimization. It is simple to implement and can be integrated readily into a variety of applications across multiple domains. In future work, we will examine AutoClip's suitability to other tasks, such as image classification, language modeling, sound event detection, and more. We will also explore applying different clipping thresholds to each layer independently, similarly to the usage of block normalized gradients in \cite{yu2017block}. Finally, we plan to investigate using moving windows, rather than the entire gradient history, as a way to reduce AutoClip's memory usage as well as make $\eta_c$ more sensitive to local rather than global training dynamics.

\bibliographystyle{IEEEtran}
\bibliography{strings,refs}

\begin{thebibliography}{10}
\providecommand{\url}[1]{#1}
\csname url@samestyle\endcsname
\providecommand{\newblock}{\relax}
\providecommand{\bibinfo}[2]{#2}
\providecommand{\BIBentrySTDinterwordspacing}{\spaceskip=0pt\relax}
\providecommand{\BIBentryALTinterwordstretchfactor}{4}
\providecommand{\BIBentryALTinterwordspacing}{\spaceskip=\fontdimen2\font plus
\BIBentryALTinterwordstretchfactor\fontdimen3\font minus
  \fontdimen4\font\relax}
\providecommand{\BIBforeignlanguage}[2]{{%
\expandafter\ifx\csname l@#1\endcsname\relax
\typeout{** WARNING: IEEEtran.bst: No hyphenation pattern has been}%
\typeout{** loaded for the language `#1'. Using the pattern for}%
\typeout{** the default language instead.}%
\else
\language=\csname l@#1\endcsname
\fi
#2}}
\providecommand{\BIBdecl}{\relax}
\BIBdecl

\bibitem{bengio1994learning}
Y.~Bengio, P.~Simard, and P.~Frasconi, ``Learning long-term dependencies with
  gradient descent is difficult,'' \emph{IEEE Transactions on Neural Networks},
  vol.~5, no.~2, pp. 157--166, 1994.

\bibitem{pascanu2013difficulty}
R.~Pascanu, T.~Mikolov, and Y.~Bengio, ``On the difficulty of training
  recurrent neural networks,'' in \emph{Proc. ICML}, 2013, pp. 1310--1318.

\bibitem{mikolov2012statistical}
T.~Mikolov, ``Statistical language models based on neural networks,'' Ph.D.
  dissertation, Brno University of Technology, 2012.

\bibitem{zhang2019gradient}
J.~Zhang, T.~He, S.~Sra, and A.~Jadbabaie, ``Why gradient clipping accelerates
  training: A theoretical justification for adaptivity,'' in \emph{Proc. ICLR},
  2019.

\bibitem{sundermeyer2012lstm}
M.~Sundermeyer, R.~Schl{\"u}ter, and H.~Ney, ``{LSTM} neural networks for
  language modeling,'' in \emph{Proc. ISCA Interspeech}, 2012.

\bibitem{luo2019convTasNet}
Y.~Luo and N.~Mesgarani, ``Conv-{T}as{N}et: Surpassing ideal time--frequency
  magnitude masking for speech separation,'' \emph{IEEE/ACM Trans. Audio,
  Speech, Language Process.}, vol.~27, no.~8, pp. 1256--1266, 2019.

\bibitem{wichern2019wham}
G.~Wichern, J.~Antognini, M.~Flynn, L.~R. Zhu, E.~McQuinn, D.~Crow, E.~Manilow,
  and J.~Le~Roux, ``{WHAM}!: Extending speech separation to noisy
  environments,'' in \emph{Proc. ISCA Interspeech}, Sep. 2019.

\bibitem{zeghidour2020wavesplit}
N.~Zeghidour and D.~Grangier, ``Wavesplit: End-to-end speech separation by
  speaker clustering,'' \emph{arXiv preprint arXiv:2002.08933}, 2020.

\bibitem{Wang2018ICASSP04Alternative}
Z.-Q. Wang, J.~{Le Roux}, and J.~R. Hershey, ``Alternative objective functions
  for deep clustering,'' in \emph{Proc. IEEE ICASSP}, Apr. 2018.

\bibitem{Kingma2015Adam}
D.~P. Kingma and J.~Ba, ``{Adam: A Method for Stochastic Optimization},'' in
  \emph{Proc. ICLR}, 2015.

\bibitem{leroux2019phasebook}
J.~Le~Roux, G.~Wichern, S.~Watanabe, A.~Sarroff, and J.~R. Hershey, ``Phasebook
  and friends: Leveraging discrete representations for source separation,''
  \emph{IEEE Journal of Selected Topics in Signal Processing}, 2019.

\bibitem{LeRoux2018SISDR}
J.~Le~Roux, S.~T. Wisdom, H.~Erdogan, and J.~R. Hershey, ``{SDR} -- half-baked
  or well done?'' in \emph{Proc. IEEE ICASSP}, May 2019.

\bibitem{huber2004robust}
P.~J. Huber, \emph{Robust statistics}.\hskip 1em plus 0.5em minus 0.4em\relax
  John Wiley \& Sons, 2004, vol. 523.

\bibitem{duchi2011adaptive}
J.~Duchi, E.~Hazan, and Y.~Singer, ``Adaptive subgradient methods for online
  learning and stochastic optimization,'' \emph{Journal of machine learning
  research}, vol.~12, no. Jul, pp. 2121--2159, 2011.

\bibitem{zeiler2012adadelta}
M.~D. Zeiler, ``Adadelta: an adaptive learning rate method,'' \emph{arXiv
  preprint arXiv:1212.5701}, 2012.

\bibitem{Tieleman2012}
T.~Tieleman and G.~Hinton, ``{Lecture 6.5---RmsProp: Divide the gradient by a
  running average of its recent magnitude},'' COURSERA: Neural Networks for
  Machine Learning, 2012.

\bibitem{yu2017block}
A.~W. Yu, L.~Huang, Q.~Lin, R.~Salakhutdinov, and J.~Carbonell,
  ``Block-normalized gradient method: An empirical study for training deep
  neural network,'' \emph{arXiv preprint arXiv:1707.04822}, Apr. 2018.

\bibitem{Hershey2016}
J.~R. Hershey, Z.~Chen, and J.~Le~Roux, ``Deep clustering: Discriminative
  embeddings for segmentation and separation,'' in \emph{Proc. IEEE ICASSP},
  Mar. 2016, pp. 31--35.

\bibitem{Luo2017}
Y.~Luo, Z.~Chen, J.~R. Hershey, J.~{Le Roux}, and N.~Mesgarani, ``Deep
  clustering and conventional networks for music separation: Stronger
  together,'' in \emph{Proc. IEEE ICASSP}, Mar. 2017, pp. 61--65.

\bibitem{kavalerov2019universal}
I.~Kavalerov, S.~Wisdom, H.~Erdogan, B.~Patton, K.~Wilson, J.~Le~Roux, and
  J.~R. Hershey, ``Universal sound separation,'' in \emph{Proc. IEEE WASPAA},
  2019, pp. 175--179.

\bibitem{Erdogan2015}
H.~Erdogan, J.~R. Hershey, S.~Watanabe, and J.~{Le Roux}, ``Phase-sensitive and
  recognition-boosted speech separation using deep recurrent neural networks,''
  in \emph{Proc. IEEE ICASSP}, Apr. 2015, pp. 708--712.

\bibitem{Kolbaek2017}
M.~Kolb{\ae}k, D.~Yu, Z.-H. Tan, and J.~Jensen, ``Multi-talker speech
  separation with utterance-level permutation invariant training of deep
  recurrent neural networks,'' \emph{IEEE/ACM Trans. Audio, Speech, Language
  Process.}, pp. 1901--1913, 2017.

\bibitem{santurkar2018does}
S.~Santurkar, D.~Tsipras, A.~Ilyas, and A.~Madry, ``How does batch
  normalization help optimization?'' in \emph{Proc. NeurIPS}, 2018, pp.
  2483--2493.

\bibitem{liu2019radam}
L.~Liu, H.~Jiang, P.~He, W.~Chen, X.~Liu, J.~Gao, and J.~Han, ``On the variance
  of the adaptive learning rate and beyond,'' in \emph{Proc. ICLR}, April 2020.

\bibitem{keskar2016large}
N.~S. Keskar, D.~Mudigere, J.~Nocedal, M.~Smelyanskiy, and P.~T.~P. Tang, ``On
  large-batch training for deep learning: Generalization gap and sharp
  minima,'' in \emph{Proc. ICLR}, 2017.

\bibitem{hochreiter1997flat}
S.~Hochreiter and J.~Schmidhuber, ``Flat minima,'' \emph{Neural Computation},
  vol.~9, no.~1, pp. 1--42, 1997.

\end{thebibliography}

\end{document}